\documentclass[aps,prl,twocolumn,superscriptaddress,amsfont,graphicx,preprintnumbers]{revtex4-1}
\renewcommand{\section}[1]{\textbf{\textit{#1.---\!}}}

\usepackage[pdftex]{graphicx}

\usepackage[colorlinks=true
,urlcolor=blue
,anchorcolor=blue
,citecolor=blue
,filecolor=blue
,linkcolor=blue
,menucolor=blue
,pagecolor=blue
]{hyperref}
\usepackage[all]{hypcap}

\pdfoutput=1
\usepackage{amsmath,amssymb}
\usepackage[usenames,dvipsnames]{color}
\usepackage{slashed} 
\usepackage{hyperref}

\makeatletter
\def\endfmffile{%
  \fmfcmd{\p@rcent\space the end.^^J%
          end.^^J%
          endinput;}%
  \if@fmfio
    \immediate\closeout\@outfmf
  \fi
  \IfFileExists{\thefmffile.mp}{\immediate\write18{mpost \thefmffile}}{}
  \let\thefmffile\relax
}
\makeatother
\usepackage[normalem]{ulem}
\newcommand{\comment}[1]{}

\newcommand{\met}{\ensuremath{\slashed{E}_T}}

\allowdisplaybreaks

\begin{document}
\title{Current LHC Constraints on Minimal Universal Extra Dimensions}
\preprint{CTPU-17-02}
\preprint{LYCEN 2017-01}

\author{Nicolas Deutschmann}
\email[]{n.deutschmann@ipnl.in2p3.fr}
\affiliation{Univ Lyon, Universite Lyon 1, CNRS/IN2P3, IPNL, F-69622, Villeurbanne, France and Centre for Cosmology, Particle Physics and Phenomenology (CP3), Universite catholique de Louvain, Chemin du Cyclotron 2, B-1348 Louvain-la-Neuve, Belgium}

\author{Thomas Flacke}
\email[]{flacke@ibs.re.kr}
\affiliation{Center for Theoretical Physics of the Universe, Institute for Basic Science (IBS),
   Daejeon, 34051, Korea}

\author{Jong Soo Kim}
\email[]{jongsoo.kim@tu-dortmund.de}
\affiliation{Center for Theoretical Physics of the Universe, Institute for Basic Science (IBS),
   Daejeon, 34051, Korea}

\begin{abstract}
In this letter, we present LHC limits on the minimal universal extra dimension (MUED) model from LHC Run 1 data and current limits from searches of the ongoing Run 2. Typical collider signals of the Kaluza-Klein (KK) states mimic generic degenerate supersymmetry (SUSY) missing transverse momentum signatures since the KK particles cascade decay into jets, leptons and the lightest KK particle which is stable due to KK parity and evades detection. We test the parameter space against a large number of supersymmetry based missing energy searches implemented in the public code {\tt CheckMATE}. We demonstrate the complementarity of employing various searches which target a large number of final state signatures, and we derive the most up to date limits on the MUED parameter space from 13 TeV SUSY searches.
\end{abstract}

\maketitle

\section{Universal Extra Dimensions -- Introduction and Review}\label{sec:intro}
Models with universal extra dimensions (UED) \cite{Appelquist:2000nn} represent a simple extension of the Standard Model which include a dark matter candidate and are testable at the LHC.\footnote{{\it C.f.} Refs.\cite{Antoniadis:1990ew,Antoniadis:1998ig} for earlier proposals of TeV scale extra dimensions. For Reviews on UED models and their phenomenology {\it c.f.} \cite{Hooper:2007qk,Servant:2014lqa}.}
The extra-dimensions are universal in the sense that all Standard Model (SM) fields are promoted to fields which propagate on the full space-time $\mathcal{M}\times X$, where $\mathcal{M}$ is the flat four dimensional (4D) Minkowski space and $X$ is a compact space.
As $X$ is compact, the momenta along the extra-dimensions are discretized. In the 4D effective theory, each extra-dimensional field yields a 4D field without extra-dimensional momentum (the zero-mode which is to be identified with the 4D SM field) as well as a Kaluza-Klein (KK) tower of excitations which are heavy partner states with the same quantum numbers as the zero-mode. The KK   mass spectrum is determined by the inverse size and the geometry of the extra-dimensions.
In the simplest case of only one extra dimension, -- which we focus on in this letter -- compactification on the orbifold $X = S^1 / Z_2$ allows to have chiral zero-modes of fermions and $A_\mu$ zero-mode for gauge fields without an additional $A_5$ mode.\footnote{For more than one extra-dimension, the compact space is not unique. {\it C.f.} Ref.\cite{Nilse:2006jv} for a classification of flat 2D orbifolds, Ref. \cite{Dobrescu:2004zi,Cacciapaglia:2009pa} for realizations. Models with spherical orbifolds have also been studied \cite{Cacciapaglia:2016xty,Maru:2009wu}.}

The 5D UED model appears to be a very simple and predictive model as it seems to have only one parameter beyond the Standard Model (BSM), the compactification radius $R$. However, as a 5D theory, the model is inherently non-renormalizable and can only be considered as an effective theory, valid below a cutoff scale $\Lambda$, which introduces an additional parameter into the model. Naive dimensional analysis \cite{Barbieri:1999tm,Papucci:2004ip,Chacko:1999hg,Bhattacharyya:2006ym} and bounds from unitarity violation in gluon KK mode scattering \cite{SekharChivukula:2001hz} suggest that the cutoff is rather low: $\Lambda R \lesssim \mathcal{O}(10 - 50)$. As a consequence, higher-dimensional operators at the cutoff scale can be phenomenologically relevant.\footnote{The least irrelevant operators are boundary localized kinetic terms and other SM-like operators. They are induced by renormalization group running and thus generically present. Their inclusion yields non-minimal UED models \cite{Flacke:2008ne} with a much larger parameter space. Another UED extension -- split UED \cite{Park:2009cs} -- includes fermion bulk mass terms. Fermion bulk mass terms are not radiatively induced and could thus be absent consistently.} In the 5D minimal UED (MUED) model \cite{Cheng:2002iz}, all higher-dimensional operators are assumed to be absent at the cutoff scale $\Lambda$, and they are only induced at lower energies due to renormalization group running, thus keeping the model a simple BSM scenario with only two parameters: the inverse compactification radius $R^{-1}$ which sets the mass scale of the first KK excitations, {\it i.e.}  of the lightest partners of the SM fields, and $\Lambda R$, which controls the number of KK modes present in the spectrum below the cutoff, and determines how much the KK mode masses and couplings are effected by one-loop running.

The phenomenology of the MUED model resembles the phenomenology of the minimal supersymmetric standard model (MSSM) in many ways. Each SM particle is accompanied by a partner particle at the first KK mode level (but in the case of UED, the partners have the same spin as the SM particle). Also, the MUED model possesses a geometric parity (``KK parity''), which is respected by loop corrections and corresponds to the reflection of the orbifold $S^1/Z_2$ at its midpoint.\footnote{In UED extensions with boundary terms and bulk fermion masses \cite{Flacke:2013pla}, KK parity is conserved if boundary terms are chosen symmetric on both boundaries and if bulk fermion masses are chosen KK parity odd.}  The lightest partner state (which in MUED is the partner of the $U(1)_Y$ gauge boson \cite{Cheng:2002iz})
represents a dark matter candidate \cite{Servant:2002aq, Kong:2005hn, Burnell:2005hm, Belanger:2010yx} which reproduces the observed dark matter relic density if $1.25 \mbox{ TeV} \lesssim R^{-1} \lesssim 1.5 \mbox{ TeV}$ \cite{Belanger:2010yx}, while for larger $R^{-1}$, the Universe would be over-closed. Via loop corrections, the KK resonances contribute to electroweak precision observables \cite{Appelquist:2000nn,Appelquist:2002wb,Baak:2011ze} and flavor physics \cite{Buras:2002ej,Buras:2003mk,Haisch:2007vb} which impose a bound of $R^{-1} \gtrsim 750 \mbox{ GeV}$ and $R^{-1} > 600 \mbox{ GeV}$.\footnote{Bounds from other precision observables, like the muon $g-2$ \cite{Nath:1999aa,Agashe:2001ra}, $Zb\bar{b}$ \cite{Oliver:2002up},  or modifications of Higgs couplings \cite{Dey:2013cqa,Kakuda:2013kba,Flacke:2013nta} are weaker.}

At the LHC, two main signal classes (non-SUSY-like and SUSY-like) allow to test UED models. MUED predicts the existence of a whole tower of partner states of which the first KK level states are KK parity odd, while the second KK level states are the lightest KK parity even BSM states. As loop-induced couplings violate KK number (whilst conserving KK parity), second KK mode states can be resonantly produced at the LHC and searched for in $Z',W'$ and colored resonance searches \cite{Datta:2005zs,Flacke:2012ke,Chang:2012wp}. The currently strongest known bound on MUED from this signal class amounts to a bound on the second KK photon mass of $m_{A^{(2)}}\gtrsim 1.4$~TeV, which corresponds to $R^{-1}\gtrsim 715$~GeV \cite{Edelhauser:2013lia} and was obtained from a recast of the CMS search for a di-lepton resonance at 8 TeV with 20.6 fb$^{-1}$ \cite{CMS:2013qca}.\\
Complementarily, SUSY searches also provide a high discovery- and exclusion potential for the MUED model. The first KK mode partners are produced in pairs and then cascade decay to the lightest state at the first KK level which itself only leaves a missing energy / momentum signature \cite{Appelquist:2000nn,Cheng:2002ab}. ATLAS and CMS performed many SUSY searches at 8 TeV and 13 TeV which are suitable to constrain the MUED parameter space. The ATLAS searches Refs.~\cite{Aad:2015mia,ATLAS-CONF-2013-062} provide explicit MUED limits, and \cite{Aad:2015mia} reports the currently strongest bound of $R^{-1}\gtrsim 900 - 950$~GeV (depending on the value of $\Lambda R$).%

In this letter, we provide the MUED bounds obtained from recasts of a large number of SUSY searches at 8 and 13 TeV, performed with {\tt CheckMATE}.

\bigskip

\section{Constraining MUED with existing SUSY searches}\label{sec:searches}
\begin{figure}[tbh]
\centering
\includegraphics[scale=0.49]{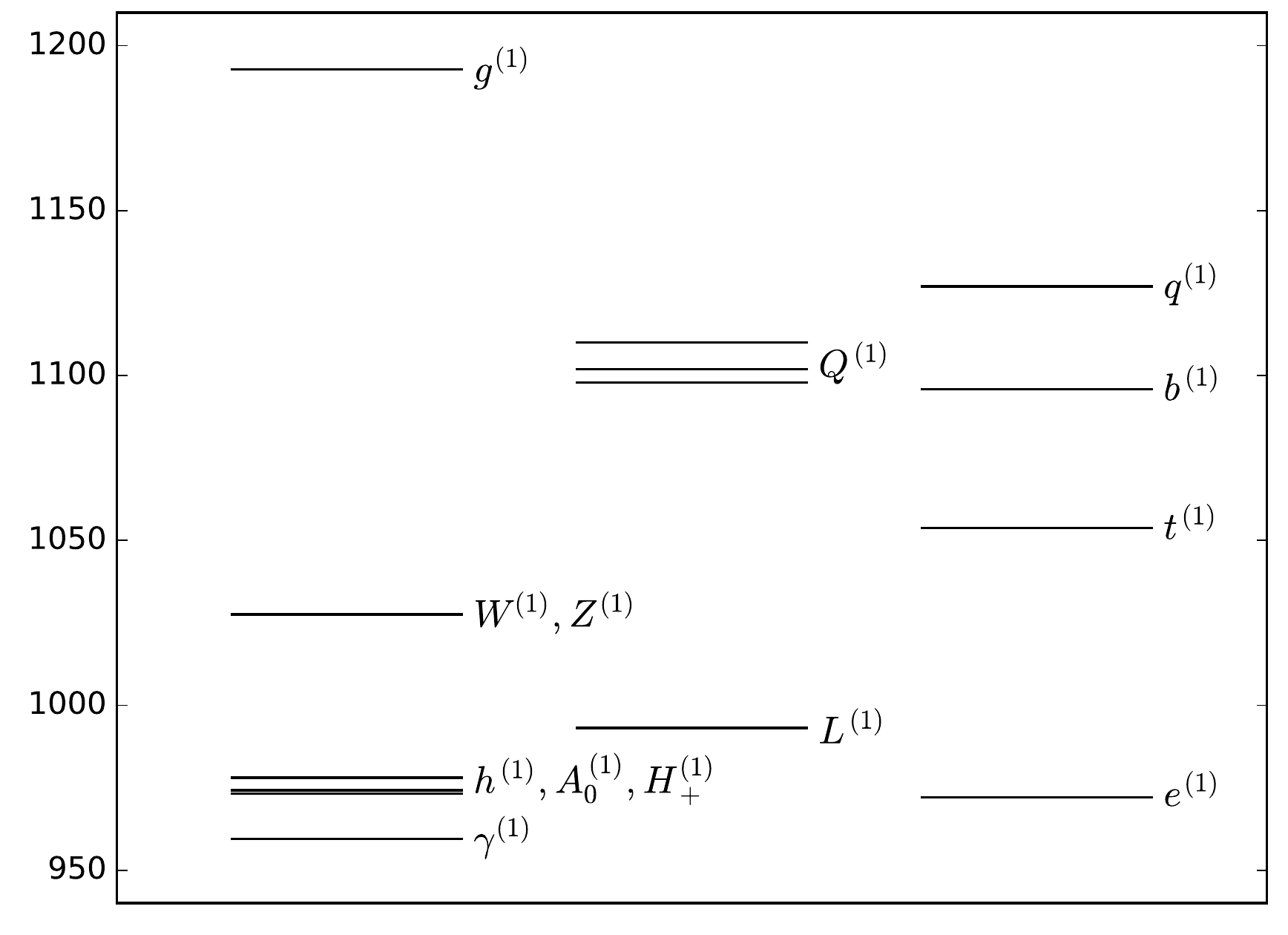}
\caption{Loop corrected KK mass spectrum in MUED for $R^{-1}$=960 GeV, $\Lambda\, R$=30 and $m_h$=125 GeV. }
\label{fig:mass_spectrum}
\end{figure}

We focus on the LHC phenomenology of the SM and the first KK level excitations. The KK zero mode level contains precisely the SM field content, including one Higgs doublet. At the first KK level, each chiral SM fermion $(Q_i,u_i,d_i,L_i,e_i)$ ($i$ is the SM family index) has one a Dirac-fermion partner $(Q^{(1)}_i,u^{(1)}_i,d^{(1)}_i,L^{(1)}_i,e^{(1)}_i)$, each SM gauge boson $g_\mu,W_\mu,B_\mu$ has a massive gauge boson partner $g^{(1)}_\mu,W^{(1)}_\mu,B^{(1)}_\mu$, and the Higgs partner sector contains a scalar, a pseudo-scalar and a charged partner  ($h^{(1)}, A^{(1)}_0, H^{(1)}_{\pm}$).\footnote{The KK gauge bosons acquire their masses dominantly from the scalar KK modes $g^{(1)}_5,W^{(1)}_5,B^{(1)}_5$, which weakly mix with the first KK Higgs modes due to electroweak symmetry breaking. Thus at each KK level, five scalar degrees of freedom are ``eaten'' while four physical scalar degrees of remain in the electroweak sector.}   The masses of the SM partners are at tree-level given by $m_n=\sqrt{(n/R)^2+ m^2_{SM} }$, suggesting a very compressed mass spectrum. However, at loop level, the near-mass-degeneracy is partially lifted (with increasing splittings for increased $\Lambda R$), making the KK gluon the heaviest state and the KK partner of the $U(1)_Y$ gauge boson the lightest state at each KK level \cite{Cheng:2002iz,Cheng:2002ab}.
A sample spectrum of the first KK resonances is shown in Fig.\ref{fig:mass_spectrum}.\footnote{{\it C.f.} Refs.\cite{Cheng:2002iz,Cheng:2002ab} for full mass expressions.} As can be seen, the mass differences can reach up to several hundreds of GeV.

In this letter, we restrict ourselves to the strong production of colored KK modes such as KK gluons and KK quarks,
\begin{equation}
pp\rightarrow g^{(1)} g^{(1)},\quad pp\rightarrow \mathcal{Q}^{(1)}_i \mathcal{Q}^{(1)}_j,\quad pp\rightarrow g^{(1)} \mathcal{Q}^{(1)}_j ,
\end{equation}
where $\mathcal{Q}^{(1)}=Q^{(1)},q^{(1)}$ and $Q^{(1)}$ denotes the $SU(2)$ doublet quark partners (or their anti-particles) and $q^{(1)} = u^{(1)},d^{(1)}$ are the $SU(2)$ singlet quark partners (or their anti-particles) taking into account all three generations.
The total production cross section is fully determined via QCD interactions and varies between $1$(8) pb and $20$(480) fb for $R^{-1}=800$ GeV and $R^{-1}=1200$ GeV with $\Lambda\,R=40$ at 8 (13) TeV, respectively.\footnote{We clearly see that the cross section increase from 8 to 13 TeV is quite significant for heavy KK masses and thus we expect a large increase in sensitivity from 8 to 13 TeV.}

The decay chain can become very complicated and in the following we briefly discuss typical decay modes of the KK states.\footnote{{\it C.f.} Refs.\cite{Cheng:2002iz,Cheng:2002ab} for expressions of the relevant zero- and first KK mode couplings at one-loop level, and an illustration of possible decay modes.} The KK gluon is the heaviest particle, and it decays into KK doublet and singlet quarks with roughly equal branching ratio.
The KK quark decay modes mainly depend on its $SU(2)$  charge. The $SU(2)$  singlet KK quark directly decays to the $U(1)_Y$ KK gauge boson $\gamma_1$ which is stable. The top KK mode also has sizable decays into the $W^{(1)}_\pm$ and $H^{(1)}_\pm$ as the KK top SU(2) doublet and singlet mix. The $SU(2)$  doublet KK quarks mainly decay into the $SU(2)$  KK gauge bosons $W^{(1)}_\pm$ and $Z^{(1)}$. The KK $W$ and $Z$ bosons are lighter than the KK quarks and thus mostly decay into leptonic KK states which themselves decay further into the lightest stable KK particle $\gamma^{(1)}$.  As a result, typical events have a relatively large lepton multiplicity, multiple jets and missing transverse momentum in the final state configuration although all decay products will be relatively soft due to the relatively compressed spectrum.

The masses and decay branching ratios are calculated with the multi purpose Monte Carlo (MC) event generator {\tt Herwig++2.7.1} \cite{Bellm:2013hwb} which is also used to generate the fully hadronized MC events employing the default parton distribution function (PDF) set MRST \cite{Martin:2002dr}. We work with the default settings in all collider tools. We feed the truth level events to {\tt CheckMATE2.0.2} \cite{Dercks:2016npn,Kim:2015wza,Drees:2013wra} which is based on the fast detector simulation {\tt Delphes 3.4.0} \cite{deFavereau:2013fsa}. {\tt Fastjet3.2.1} is used for the jet clustering \cite{Cacciari:2011ma}. {\tt CheckMATE} allows for easy testing of whether a model point is excluded or not at $95\%$ confidence level against current ATLAS and CMS searches at the LHC. {\tt CheckMATE} requires an event file and the corresponding production cross section as input. For the study presented here, we take the leading order cross section from {\tt Herwig++}, which yields a very conservative estimate of the constraints as we do not re-scale the leading order estimate with a $\mathcal{K}$ factor.
In principle, MC events should be generated with at least one additional parton at matrix element level and matched with the {\tt Herwig++} parton shower since the MUED particle spectrum is relatively degenerate \cite{Dreiner:2012gx,Dreiner:2012sh,Drees:2012dd}. However, this is beyond the scope of the current work. Our simple approach will certainly introduce a non-negligible systematic uncertainty in our predictions for mass splitting smaller than 100 GeV, which in particular occurs in the low $\Lambda R\le5$ region.

We have performed a grid scan in the $R^{-1}$ -- $\Lambda\, R$ plane with the SM Higgs mass fixed to $m_h=125$ GeV. For each grid point, $10^5$ events have been generated.
We test each grid point against all ATLAS and CMS searches implemented into {\tt CheckMATE}. However, not every search is sensitive to MUED and we only show the relevant 8 and 13 TeV studies in Table \ref{tab:lhc_searches8tev} and \ref{tab:lhc_searches13tev}. Most searches are validated and the validation notes can be found on the official webpage \footnote{\url{https://checkmate.hepforge.org/}}.\footnote{The implementation of the search \cite{ATLAS-CONF-2013-062} has been only partially validated and we did not include it in our search. The search \cite{Aad:2015mia} presents limits on the MUED parameter space and we plot their limits in Fig.~\ref{fig:results_8tev}. The conference notes \cite{ATLAS-CONF-2016-054,ATLAS-CONF-2016-078} did not provide cutflows. However, both searches are based on \cite{Aaboud:2016zdn,Aad:2016qqk} with minor modifications and the two latter searches are validated.} Each analysis typically contains a large number of signal regions which target different mass hierarchies and final state multiplicities. For a given search, the {\it best signal region} for each point in the parameter space is defined by {\tt CheckMATE} as that with the largest {\it expected} exclusion potential. This criterium is then repeated to select the {\it best search} which is defined as the search whose {\it best signal region} has the largest expected limit. This means that the best observed limit is not always used but it  ensures that the result is less sensitive to downward fluctuations in the data that are bound to be present when scanning over many searches and many signal regions. Once a {\it best search} is selected for a given point in parameter space, the signal yield in our model is then compared to the observed limit at 95\% confidence level \cite{Read:2002hq},

\begin{equation}
r=\frac{S-1.96\cdot\Delta S}{S_{\rm exp}^{95}}
\label{eq:r}
\end{equation}

where $S$, $\Delta S$, and $S_{\rm exp}^{95}$ denotes the number of signal events, its theoretical uncertainty and the experimentally determined 95$\%$ confidence level limit on the number of signal events $S$, respectively.We only consider the statistical uncertainty due to the finite Monte Carlo sample with $\Delta S=\sqrt{S}$. The quantity $S-1.96\cdot\Delta S$ corresponds to the 95\% lower bound on our prediction for the number of signal event, which ensures that the limits we set are conservative~\cite{Drees:2013wra}. The $r$ value is only calculated for the expected best signal region.
{\tt CheckMATE} does not combine signal regions nor analyses in order to optimize exclusion since this would require knowledge from the experimental collaborations. We consider a model point as excluded if $r$ is larger than one. However, due to theoretical uncertainties we cannot evaluate directly such as missing next to leading
order calculations, PDF uncertainties, details of the parton shower and
the finite MC event sample, we assign a conservative uncertainty to our
signal prediction. As a result we define a point as definitively allowed
for $r < 2/3$ and definitively excluded for $r > 3/2$, which should be large
enough to account for the aforementioned effects.

\begin{table}[tbh]
\begin{center}
\begin{tabular}{l|l|l}
Reference & Final State & $\mathcal{L}$ [fb$^{-1}$]\\
\hline
1403.4853 (ATLAS) \cite{Aad:2014qaa} & 2$\ell$+\met& 20.3 \\
1404.2500 (ATLAS) \cite{Aad:2014pda} & SS 2$\ell$ or 3$\ell$ & 20.3\\
1405.7875 (ATLAS) \cite{Aad:2014wea} & jets + \met & 20.3\\
1407.0583 (ATLAS) \cite{Aad:2014kra} & 1$\ell$+($b$) jets+\met & 20.0\\
1407.0608 (ATLAS) \cite{Aad:2014nra} & monojet+\met & 20.3\\
1402.7029 (ATLAS)  \cite{Aad:2014nua} & 3$\ell$+\met& 20.3 \\
1501.03555 (ATLAS) \cite{Aad:2015mia} & 1$\ell$+jets+\met & 20.3\\
1303.2985 (CMS) \cite{Chatrchyan:2013mys} & $\alpha_{T}$+$b$ jets & 11.7\\
1405.7570 (CMS) \cite{Khachatryan:2014qwa} & 1,\,SS-OS2,\,3,\,4$\ell$+\met& 20.3 \\
\end{tabular}
\end{center}
\caption{8 TeV analyses used in our study.
Articles are
shown by their arXiv number. The middle column denotes the target final
state, and the third column shows the total integrated luminosity.}
\label{tab:lhc_searches8tev}
\end{table}

\begin{table}[tbh]
\begin{center}
\begin{tabular}{l|l|l}
Reference & Final State & $\mathcal{L}$ [fb$^{-1}$]\\
\hline
1605.03814 (ATLAS) \cite{Aaboud:2016zdn} & jets+\met & 3.2 \\
1605.04285 (ATLAS) \cite{Aad:2016qqk} & 1$\,\ell$+jets+\met &  3.2\\
ATLAS-CONF-2016-054 \cite{ATLAS-CONF-2016-054} & 1$\ell$+jets+\met& 14.8 \\
ATLAS-CONF-2016-076 \cite{ATLAS-CONF-2016-076} & 2$\ell$ (stop search) &13\\
ATLAS-CONF-2016-078 \cite{ATLAS-CONF-2016-078} & jets+\met & 13.3\\
\end{tabular}
\end{center}
\caption{13 TeV analyses used in our study. The *CONF*
  papers are only published as conference proceedings. The middle column denotes the target final state, and the third column shows the total integrated luminosity.}
\label{tab:lhc_searches13tev}
\end{table}

\bigskip


\section{Results}\label{sec:results}
\begin{figure}[t]
\centering
\includegraphics[scale=0.32]{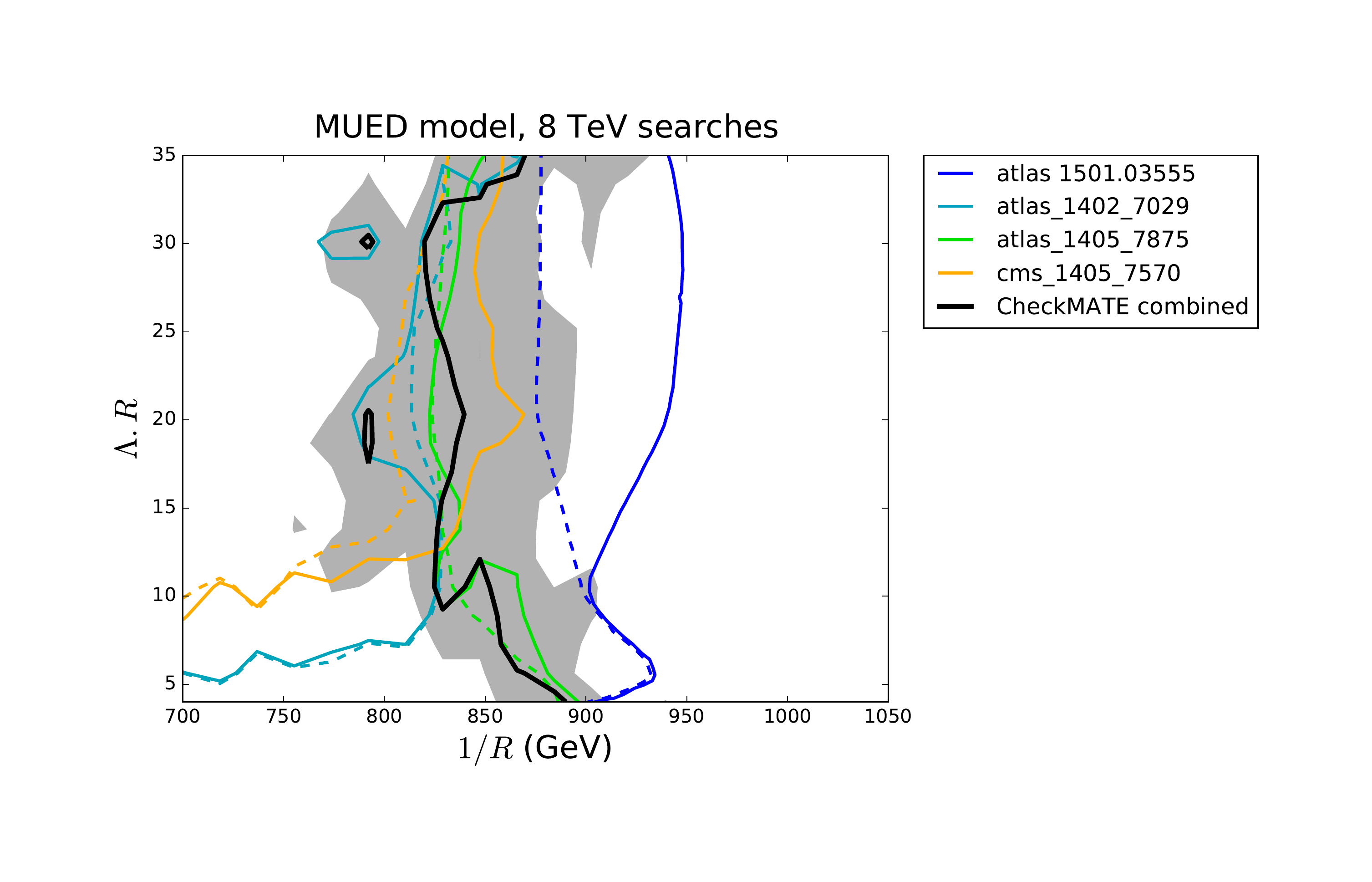}
\caption{95$\%$ confidence level limit on the $R^{-1}$ -- $\Lambda\, R$ plane from ATLAS and CMS searches performed at the center of mass energy $\sqrt{s}=8$ TeV. The solid and dashed line denotes the observed and expected 95$\%$ confidence level limit, respectively from the searches employed in our paper. The total integrated luminosity used by the various searches are summarized in Table~\ref{tab:lhc_searches8tev}. The black solid line denotes the best limit excluding Ref.~\cite{Aad:2015mia} and the shaded area corresponds to our theoretical uncertainty. We fixed the SM Higgs mass to $m_h$=125 GeV.}
\label{fig:results_8tev}
\end{figure}

\begin{figure}[t]
\centering
\includegraphics[scale=0.32]{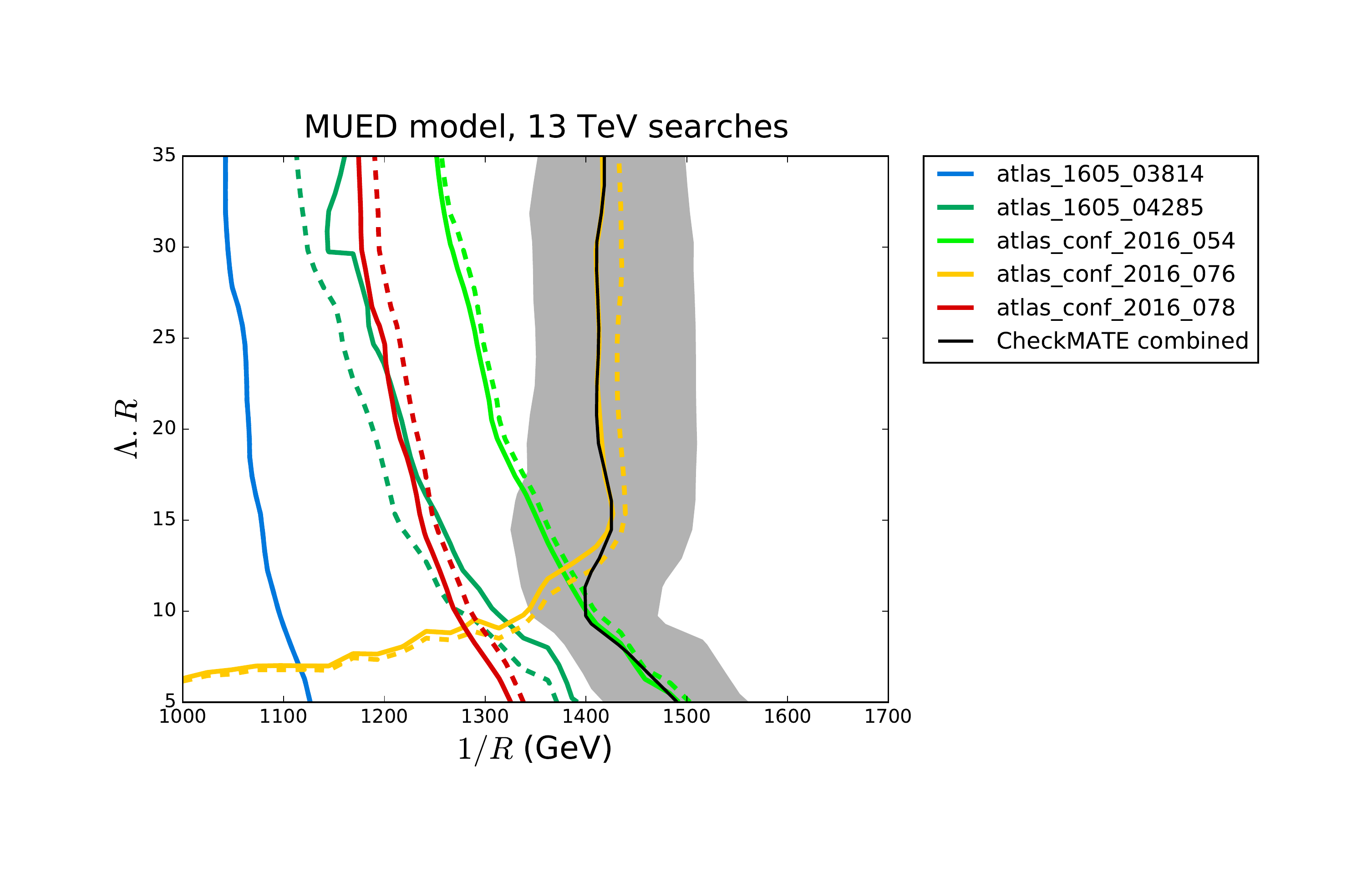}
\caption{95$\%$ confidence level limit on the $R^{-1}$ -- $\Lambda\, R$ plane from ATLAS and CMS searches performed at the center of mass energy $\sqrt{s}=13$ TeV. The solid and dashed line denotes the observed and expected 95$\%$ confidence level limit, respectively from the searches employed in our paper. The total integrated luminosity used by the various searches are summarised in Table~\ref{tab:lhc_searches13tev}. The black solid line denotes the best limit and the shaded area corresponds to our theoretical uncertainty. We fixed the SM Higgs mass to $m_h$=125 GeV.}
\label{fig:results_13tev}
\end{figure}

Our numerical results are shown in Fig.~\ref{fig:results_8tev} and \ref{fig:results_13tev}. We first give a brief summary of the 8 TeV results. Then we present the most up to date limits on the two dimensional MUED parameter space from 13 TeV data.

Run 1 was very successful and both LHC experiments, ATLAS and CMS published a large number of searches sensitive to a multitude of final state signatures. We want to investigate the impact of taking into account all relevant 8 TeV searches on MUED since only a few dedicated studies have been considered for MUED so far. In Fig.~\ref{fig:results_8tev}, we fix the SM Higgs mass to 125 GeV and we present the excluded region at 95 $\%$ confidence level in the $R^{-1}$ -- $\Lambda\, R$ plane.

Ref.~\cite{Aad:2015mia} has an explicit MUED search included, and we show their results for reference. As can be seen, the search Ref.~\cite{Aad:2015mia} provides the best sensitivity to the MUED parameter space. In particular its dilepton signal regions targeting soft as well as hard leptons provide strong constraints. It should be pointed out that the observed limit of \cite{Aad:2015mia} is actually much better than the expected one due to mild down fluctuations in the observed data. As a consequence, a compactification scale up to 950 GeV for $\Lambda\,R$=30 is excluded.
Now, we want to investigate how the other SUSY searches perform in comparison and we explore the sensitivity on MUED excluding Ref.~\cite{Aad:2015mia}. The limits of other SUSY searches, as obtained with {\tt CheckMATE} are shown in green, turquoise and  brown. We only show results of searches out of the list provided in Table~\ref{tab:lhc_searches8tev} which yield the best expected bound (shown as dashed lines) on $R^{-1}$ for some value of $\Lambda R$. The recast observed bounds of the individual searches are shown as solid lines. The black solid curve denotes the 95 \% confidence level bound as determined from Eq.~(\ref{eq:r}) for $r=1$,\footnote{The apparent discrepancy between the limit shown by this line and trying to follow the individual limits is a result of the interpolation we performed to produce this figure. In particular, the displayed expected limit for the CMS search is in reality briefly the best search around $\Lambda R=20$, which is not displayed because the interpolation smoothed the curve. This explains the outward bulge of the limit in that region.} and the grey shaded area is our estimate for the uncertainty and denotes the region with $0.67<r<1.5$.
Considering searches other than Ref.~\cite{Aad:2015mia}, the limit is mainly driven by  \cite{Aad:2014wea}, \cite{Aad:2014nua}, and \cite{Khachatryan:2014qwa}. \cite{Aad:2014wea} is the 'vanilla' ATLAS multijet and large missing transverse momentum search which performs well for sizable $\Lambda\, R$. For small $\Lambda\,R\sim5$ GeV, the mass difference between the KK gluon $g_1$ and the lightest KK mode $\gamma_1$ diminishes quite rapidly with a splitting of $\sim 100$ GeV. In this regime, monojet searches can become competitive. However, since we did not generate matched event samples, the systematic uncertainty on the exclusion limit in this region of parameter space would be too large and thus the results should be interpreted with great care. Ref.  \cite{Aad:2014nua} and \cite{Khachatryan:2014qwa} are dedicated multi-lepton and missing transverse momentum searches. Multiple leptons in the final state are very handy in suppressing the SM background. It is clear that for small $\Lambda\, R$ values, the leptons become too soft and thus the multi-lepton searches loose their sensitivity very quickly. Again, dedicated studies targeting soft leptons such as Ref.~\cite{Aad:2015mia} and possibly monojet searches will improve sensitivity in this degenerate region.

To summarize, we have demonstrated that a large number of searches can provide complementary constraints on the MUED parameter space. The strongest constraints from 8 TeV searches are obtained from the dedicated MUED motivated signal region of Ref.~\cite{Aad:2015mia}, but signal regions of other SUSY searches yield bounds which are not substantially weaker. Other 8 TeV searches such as \cite{TheATLAScollaboration:2013xha}          could be sensitive to the signals we consider, but are currently not implemented or validated in {\tt CheckMATE}.

Now, we turn our attention to LHC searches at $\sqrt{s}=13$~TeV. So far, ATLAS and CMS did not present dedicated MUED results from Run 2. In Fig.~\ref{fig:results_13tev}, we display the excluded region from the ongoing Run 2 as obtained from our recast study with {\tt CheckMATE}. We include two published searches \cite{Aaboud:2016zdn,Aad:2016qqk} from the early Run 2 data and three conference notes with roughly 14 fb$^{-1}$ of collected data \cite{ATLAS-CONF-2016-078,ATLAS-CONF-2016-076,ATLAS-CONF-2016-054}. Other searches such as \cite{Aaboud:2016ejt,ATLAS:2016kjm,ATLAS:2017cjl,Khachatryan:2017qgo,Sirunyan:2016jpr,Khachatryan:2016fll,Khachatryan:2016kdk,CMS:2017vro,CMS:2017iir}          could also be sensitive to the signals we consider, but are not yet implemented or validated in {\tt CheckMATE}. We observe a clear improvement of the limits as compared to Run 1. The mass scale up to $R^{-1}=1500$ GeV can be excluded at 95$\%$ confidence level. The search \cite{ATLAS-CONF-2016-054} requiring one isolated lepton performs better than the corresponding search with a lepton veto \cite{ATLAS-CONF-2016-078}. For both searches, the limits become much weaker for increasing $\Lambda R$. This might be quite surprising since naively one would expect that for increasing mass splitting, the efficiency in the signal regions would increase. However, increasing $\Lambda\, R$ raises also the overall mass scale of the hard process which results in a large reduction of the hadronic cross section. In addition, the best signal region of Ref.~\cite{ATLAS-CONF-2016-054} is GG2J which targets relatively heavy and compressed spectra. Sensitivity is enhanced for events with relatively large jet recoils against the pair produced KK partons which results in large missing transverse momentum. However, for increasing mass splitting, the missing transverse energy is reduced and thus we observe that the best limits are derived for relatively small $\Lambda\,R$. On the other hand, the dilepton stop search \cite{ATLAS-CONF-2016-076} performs very well even for large $\Lambda\,R$ values. Here, a large mass splitting results in more energetic leptons which improves the sensitivity. However, the response degrades very quickly for very small $\Lambda\, R$ as in this region, the leptons tend to be too soft to be detected.

\bigskip

\section{Conclusions}\label{sec:conclusions}
We have determined the current limits on the MUED parameter space from LHC supersymmetry searches at 8 and 13 TeV. After the end of Run 1, a large number of supersymmetry searches targeting a vast number of final state topologies have been published. We demonstrated the complementarity of using all available searches in order to constrain MUED. Run 2 is ongoing and the number of searches is quite restricted compared to Run 1. However, the large gain in parton luminosity allows for impressive improvement in the exclusion of parameter space. The dilepton stop search performs very well, and we have derived the most stringent limits on the MUED parameter space so far with $R^{-1}\approx 1400$ GeV excluded for $\Lambda\, R\sim10$. This limit is conservative in the sense that it does not include and $\mathcal{K}$ factor. For a $\mathcal{K}$ factor of 1.5, the limit would be increased to $R^{-1}\approx 1500$ GeV. MUED yields the observed dark matter relic density if co-annihilation and second KK mode resonant (co-) annihilation are taken into account if $1.25 \mbox{ TeV} \lesssim R^{-1} \lesssim 1.5 \mbox{ TeV}$ \cite{Belanger:2010yx} while larger $R^{-1}$ predicts a too large dark matter relic density. Thus the results from ATLAS and CMS Run 2 recasts of SUSY searches presented in this letter show that the MUED model ({\it i.e.} the simplest UED benchmark model) is getting in tension with observation.

\bigskip

\section{Note added} While we were finishing this letter, we became aware of the complementary study Ref.~\cite{Beuria:2017jez}, which studies MUED bounds from SUSY searches with an MUED implementation in {\tt Pythia 6} and {\tt Pythia 8} for event generation and {\tt CheckMATE}. We thank the authors of Ref.~\cite{Beuria:2017jez} for communication and for making a preliminary draft of their
paper available to us. In the overlapping regions of the two studies, the results were found to be in reasonable agreement.

\bigskip

\acknowledgments
\section{Acknowledgments}
We thank Kyoungchul Kong and Seong Chan Park for their contribution and collaboration in the beginning of this work as well as for many helpful comments. We also thank Simon Platzer and Peter Richardson for information and details on the MUED implementation in {\tt Herwig}. The work of J.S.K and T.F. was supported by IBS under the project code, IBS-R018-D1. N.D. and T.F. acknowledge the Partenariat Hubert Curien (PHC) STAR project no.34299VE, and partial support from the CNRS LIA FKPPL.

\bibliographystyle{utphys}
\bibliography{paperbibMod}

\end{document}